\documentclass[12pt]{article}
\usepackage{epsfig}
\title{\bf Proposed Experiments to Test the Unified Description
of Gravitation and Electromagnetism through a Symmetric Metric}
\author{  Murat \"Ozer \\
Department of Physics, College of Science,
King Saud University,\\
P. O. Box 2455, Riyadh 11451, Saudi Arabia\\
E-mail: mozer@ksu.edu.sa}
\begin{document}
\maketitle
\begin{abstract}
\noindent If gravitation and electromagnetism are both described in terms
of a symmetric metric tensor, then the deflection of an electron beam
by a charged sphere should be different from its deflection according to
the Reissner-Nordstr{\o}m solution of General Relativity. If such a 
unified description is true, the equivalence principle for the electric 
field implies that the photon has a nonzero effective electric charge-to 
mass ratio 
and should be redshifted as it moves in  an electric field and be
deflected in a magnetic field. Experiments to test these predictions are 
proposed.
\end{abstract}

Of all the unification schemes for gravitation and electromagnetism 
suggested so far, the simplest is the one through a symmetric metric tensor 
$g_{\mu\nu}$ [1]. 
In this scheme gravitation and electromagnetism curve the spacetime
in exactly the same way, as a result of which the interpretation of the
metric tensor as the gravitational field proper must be given up. If this
scheme of unified description does indeed correspond to reality, it must 
possess testable deviations from Einstein's general relativity 
(hereafter GR) [2] as well as new physical phenomena.  The purpose of
this letter is, therefore, to propose experiments through which this new 
scheme can be tested. To this end, we shall discuss three topics and
their experimental implications.

{\bf I. The Line Element for a Spherically Symmetric Distribution of 
Matter and Charge:} In Einstein's GR theory, the gravitational field around
a spherical distribution of mass  $M$ and charge $Q$ located at $r=0$ is 
described by the field equation
\begin{equation} 
R^{\mu\nu}=\frac{8\pi G}{c^4}T^{\mu\nu}_{EM},
\end{equation}
where $T^{\mu\nu}_{EM}$ is the usual traceless tensor of the electromagnetic
field of the charge $Q$. The spherically symmetric solution of eq.(1) for 
the line element (the invariant interval) is known as the 
Reissner-Nordstr{\o}m solution [3,4]. It is given by
\footnote{We use the conventions of Misner, Thorne, and Wheeler [5] for
metrics, curvatures, etc.}
\begin{eqnarray} 
ds^2=-\left(1-2\frac{GM}{c^2r}+\frac{Gk_eQ^2}{c^4r^2}\right)c^2dt^2+
\left(1-2\frac{GM}{c^2r}+\frac{Gk_eQ^2}{c^4r^2}\right)^{-1}dr^2+\nonumber\\
r^2d\theta^2+r^2sin^2\theta d\phi^2,
\end{eqnarray}
where $G$ and $k_e$ are the gravitational and electric constants, c is
the speed of light. In our 
scheme, the equation describing the dynamical effects of the gravitational
as well as the electric field around such a mass and electric charge
distribution on a test particle of mass $m$ and electric charge $q$ is
\begin{equation} 
R^{\mu\nu}=0.
\end{equation}
The solution of eq.(3) is similar to the Schwarzschild solution [6] and is 
easily found to be
\begin{eqnarray} \label{eq.(4)}
ds^2=-\left(1-2\frac{GM}{c^2r}+2\frac{q}{m}\frac{k_eQ}{c^2r}\right)c^2dt^2+
\left(1-2\frac{GM}{c^2r}+2\frac{q}{m}\frac{k_eQ}{c^2r}\right)^{-1}dr^2
\nonumber\\
+r^2d\theta^2+r^2sin^2\theta d\phi^2.
\end{eqnarray}
Comparison of the third terms in  $g_{00}$ of equations (2) and (4) reveal 
the philosophy of our unification. In eq.(1), the electric field of the 
charge distribution contributes to the gravitational field of the matter. 
Whereas in our scheme, there is an equivalence principle for the 
electromagnetic field as well [1], and the right-hand side of eq.(3) is 
zero, as opposed to eq.(1) of GR; the electric field does not contribute to
the gravitational field, it asserts itself separately. To test which of the 
third terms in $g_{00}$ of equations (2) and (4) reflects the physical 
reality, consider a positively charged metallic sphere of radius $R$, mass 
$M$, and electric charge $Q$. The electric potential on the surface of the 
sphere is
\begin{equation} \label{eq.(5)}
V(R)=\frac{k_eQ}{R},
\end{equation} 
in terms of which the $g_{00}$ are
\begin{eqnarray} \label{eq.(6)}
g^{RN}_{00}= -\left(1-2\frac{m_G}{r}+\frac{GR^2}{k_ec^4r^2}V(R)^2\right);
\nonumber \\
g^{MGR}_{00}= -\left(1-2\frac{m_G}{r}+2\frac{q}{m}\frac{R}{c^2r}V(R)\right),
\end{eqnarray}
where the first one corresponds to the Reissner-Nordstr{\o}m (RN) solution 
and
the second one to ours, which we call ``modified general relativity'' 
(hereafter MGR), and $m_G=GM/c^2$. 
Now, for a sphere of $M=1kg$, $R=5cm$, and  
an electric potential of $10^3V$ on the surface of the sphere, we have, 
for an electron just grazing the sphere
\footnote{Note that in the Reissner-Nordstr{\o}m case, the contribution of 
the electric charge of the sphere
to its gravitational field turns out to be much smaller than the mass term
$2GM/c^2r$ for reasonable values of $r$.}
\begin{eqnarray} \label{eq.(7)}
g^{RN}_{00}= -\left(1-1.48\times 10^{-26}+9.19\times 10^{-49}\right)
\approx -1; \nonumber \\
g^{MGR}_{00}= -\left(1-1.48\times 10^{-26}-3.91\times  10^{-3}\right)
\approx -0.996.
\end{eqnarray}
Thus, the space around such a charged sphere is extremely closed to being 
flat in the Reissner-Nordstr{\o}m case and is approximated perfectly by the
metric of special relativity, the Minkowski metric. In our case, however,
there is a great deal of deviation from flatness that can assert itself in
the trajectory of an electron moving in the viscinity of the sphere. The
trajectory of an electron ($q=-e$) moving in the gravitational and electric
fields, however weak they are, of such a sphere is described by
\begin{equation}\label{eq.(8)}
\frac{d^2x^{\mu}}{ds^2}+\Gamma^{\mu}_{\alpha\beta}\frac{dx^{\alpha}}{ds}
\frac{dx^{\beta}}{ds}=\frac{q}{mc^2}F^{\mu}_{\alpha}\frac{dx^{\alpha}}{ds},
\end{equation}
in the Reissner-Nordstr{\o}m case with $\Gamma^{\mu}_{\alpha\beta}$
calculated from eq.(2), and by
\begin{equation}\label{eq.(9)}
\frac{d^2x^{\mu}}{ds^2}+\Gamma^{\mu}_{\alpha\beta}\frac{dx^{\alpha}}{ds}
\frac{dx^{\beta}}{ds}=0,
\end{equation}
in our scheme with $\Gamma^{\mu}_{\alpha\beta}$ calculted from eq(4).

To simplify the notation, let us, as usual, write the line element in the
form
\begin{equation}\label{eq.(10)}
ds^2=-e^{\eta}c^2dt^2+e^{-\eta}dr^2+r^2d\theta^2+r^2sin^2\theta d\phi^2.
\end{equation}
The nonzero components of
\begin{equation}\label{eq.(11)}
\Gamma^{\mu}_{\alpha\beta}=\frac{1}{2}g^{\mu\nu}\left(g_{\nu\alpha,\beta}+
g_{\nu\beta,\alpha}-g_{\alpha\beta,\nu}\right),
\end{equation}
that we need in our calculation are
\footnote{The  other nonzero components of $\Gamma^{\mu}_{\alpha\beta}$
that are not required in our calculation have not been quoted here.}
\begin{eqnarray}\label{eq.(12)}
\Gamma^0_{01}&=& \Gamma^0_{10}=\frac{1}{2}\frac{d\eta}{dr},\hspace{0.5cm}
\Gamma^3_{13}=\Gamma^3_{31}=\frac{1}{r},
\end{eqnarray}
Using
\begin{equation} \label{eq.(13)}
A_{\mu}=\left(-\Phi_E,\vec A\right)=\left(-k_e\frac{Q}{r},0\right),
\end{equation}
the nonzero components of the electromagnetic field strength tensor
\begin{equation} \label{eq.(14)}
F_{\mu\nu}=\frac{\partial A_{\nu}}{\partial x^{\mu}}-
\frac{\partial A_{\mu}}{\partial x^{\nu}},
\end{equation}
are
\begin{equation} \label{eq.(15)}
F_{01}=-F_{10}=-k_e\frac{Q}{r^2}.
\end{equation}
Confining the motion of the electron in the $\theta=\pi /2$ plane not only 
simplifies the calculation a lot but also the experiment to be described
later. We, then, obtain the following euations from eq.(8) for the 
coordinates $x^0=ct$ and $x^3=\phi$
\begin{equation} \label{eq.(16)}
\frac{d^2t}{ds^2}+\frac{d\eta}{dr}\frac{dr}{ds}\frac{dt}{ds}=\frac{q}{mc^3}
e^{-\eta}\frac{k_eQ}{r^2}\frac{dr}{ds},
\end{equation}
\begin{equation} \label{eq.(17)}
\frac{d^2\phi}{ds^2}+\frac{2}{r}\frac{dr}{ds}\frac{d\phi}{ds}=0,
\end{equation}
where we have put $d\theta/ds=0$. A further simplification is achieved by
trading the equation for the coordinate $x^1=r$ with the one that follows
from the condition of timelike geodesics
\begin{equation} \label{eq.(18)}
g_{\mu\nu}\frac{dx^{\mu}}{ds}\frac{dx^{\nu}}{ds}=-1,
\end{equation}
which gives
\begin{equation} \label{eq.(19)}
e^{-\eta}\left(\frac{dr}{ds}\right)^2+r^2\left(\frac{d\phi}{ds}\right)^2
-e^{\eta}c^2\left(\frac{dt}{ds}\right)^2+1=0.
\end{equation}
Equations (16) and (17) can be integrated to yield, respectively
\begin{equation} \label{eq.(20)}
\frac{dt}{ds}=\frac{e^{-\eta}}{c}\left(-\frac{qk_eQ}{mc^2}\frac{1}{r}
+a\right),
\end{equation}
\begin{equation} \label{eq.(21)}
r^2\frac{d\phi}{ds}=h,
\end{equation}
where $a$ and $h$ are integration constants. Noting that $dr/ds=(dr/d\phi)
(d\phi/ds)$ and inserting equations (20) and (21) in eq.(19) and then 
dividing by $e^{-\eta}$ we get
\begin{equation} \label{eq.(22)}
\left(\frac{du}{d\phi}\right)^2+u^2e^{\eta}-\left(-\frac{qk_eQ}{mc^2}u
+a\right)^2\frac{1}{h^2}+\frac{e^{\eta}}{h^2}=0,
\end{equation}
where we have set $u=1/r$. For the Reissner-Nordstr{\o}m solution we now
put $e^{\eta} \approx 1$. Differentiating eq.(22) with respect to $\phi$
and removing the factor $du/d\phi$ we get
\begin{equation} \label{eq.(23)}
\frac{d^2u}{d\phi^2}+u=\frac{m_E}{h^2}+\frac{m_E^2}{h^2}u
\end{equation}
where we have set the constant $a=1$ so that when $h=l/mc$, with 
$l=mr^2\dot \phi$ being the ordinary angular momentum, the first term on the 
right-hand side of 
eq.(23) agrees with the Newtonian (hereafter N) expression
\begin{equation} \label{eq.(24)}
\frac{d^2u}{d\phi^2}+u=\frac{m^2c^2}{l^2}m_E.
\end{equation}
Here 
\begin{equation} \label{eq.(25)}
m_E=-\frac{q}{m}\frac{k_eQ}{c^2}=-\frac{q}{mc^2}RV(R)
\end{equation}
has the dimension of length and corresponds to  $m_G=GM/c^2$ in the
Schwarzschild solution. Eq.(23)  describes the trajectory of a charged 
test particle when $g_{11}\approx -g_{00}\approx 1$ in the 
Reissner-Nordstr{\o}m solution. Hence, it also describes exactly the 
trajectory of a test charge
in an electric field according to special relativity. The second term on
the right-hand side of eq.(23) is a special relativistic correction to
the Newtonian result.

As for eq.(9), we get
\begin{equation} \label{eq.(26)}
\frac{d^2t}{ds^2}+\frac{d\eta}{ds}\frac{dt}{ds}=0
\end{equation}
instead of eq.(16), and 
\begin{equation} \label{eq.(27)}
\frac{dt}{ds}=\frac{e^{-\eta}}{c}
\end{equation}
instead of eq.(20) with $a=1$. Equations (17) and (21) do not change. 
Proceeding as before, we find
\begin{equation} \label{eq.(28)}
\frac{d^2u}{d\phi^2}+u=\frac{m_E}{h^2}+3m_Eu^2.
\end{equation}
Recall that terms involving $m_G$ on the right-hand sides of equations
(23) and (28) have been dropped because $m_G<<m_E$ for the metallic sphere
we are considering. It should be noted that when $m_E$ is replaced with 
$m_G$ in eq.(28), the equation of a neutral test particle of mass $m$ moving
in the Schwarzschild field of a spherical mass $M$ is obtained. Since 
$mch$ is the conserved angular momentum of the
test charge in its rest frame, we need to express $h$ in terms of $l$, the
ordinary angular momentum of the test charge in the laboratory frame (with
respect to the coordinate time  $t$). In the \mbox{Reissner-Nordstr{\o}m} 
case,equations (20) and (21), with $e^{-\eta}=1$, yield
\begin{equation} \label{eq.(29)}
h=\frac{l}{mc}\left(1+m_Eu\right),
\end{equation}
 and in our scheme equations (21) 
and (27), with $e^{-\eta}=\left(1-2m_Eu\right)^{-1}$,  yield
\begin{equation} \label{eq.(30)}
h=\frac{l}{mc}\left(1-2m_Eu\right)^{-1}.
\end{equation}
 Equations (23) and (28) then 
reduce to  
\begin{equation} \label{eq.(31)}
\frac{d^2u}{d\phi^2}+u=\frac{m^2c^2}{l^2}\frac{m_E}{\left(1+m_Eu\right)},
\end{equation}
which is the orbit equation for the Reissner-Nordstr{\o}m solution, and
\begin{equation} \label{eq.(32)}
\frac{d^2u}{d\phi^2}+u=\frac{m^2c^2}{l^2}m_E\left(1-2m_Eu\right)^2+
3m_Eu^2,
\end{equation}
which is the orbit equation in our scheme.

We now propose the following experiment to distinguish between the two
equations; (31) and (32): Consider a vacuum chamber in the shape of a 
rectangular metallic box. Let a metallic ball of radius $R$ positively
charged to a potential of $V(R)$ be hanged freely from an insulating thread.
Let an electron gun be located at angle $\alpha$ on the equatorial plane
of the ball at a distance $r_i$ from the ball's center. The point of
emergence of the electrons may be taken to be on the negative $y$ axis and
thus has $\phi=3\pi/2$. Put a calibrated phosphorous screen on
the positive y axis at $\phi=5\pi/2$. Make a large enough glass 
window on the side of the box facing the screen (or monitor the position
of the electron beam on the screen electronically) to observe where the
electron beam hits on the screen. Equations (31) and (32) can be solved
numerically for $u$, and hence for $r$. The two initial conditions 
required are $u(\phi=3\pi/2)=r^{-1}_i$ and 
\mbox{$du/d\phi(\phi=3\pi/2)=\sqrt{1-sin^2\alpha}/(r_isin\alpha)$}, where 
as above $\alpha$ is the
angle the initial velocity $v_i$ of the electrons makes with the positive
y axis. In obtaining the second initial condition we have made use of
\mbox{$dr/dt=\dot r=(dr/d\phi)(d\phi/dt)=r'\dot \phi$}, 
\mbox{$v^2=\dot r^2+r^2\dot \phi^2$}, and $l=mr^2\dot \phi=mv_ib$, where
$b=r_isin\alpha$ is the impact parameter of the electrons. The solutions of
the equations (31) and (32) can thus be found numerically at any value
of the angle $\phi$, and especially on the positive $y$ axis.
 We have tabulated some examplary results in Tables
1 and 2
\footnote{In our calculations we have used the 
relativistic expression $eV_{AC}=m_ec^2/\sqrt{1-v_i^2/c^2}-m_ec^2$ to 
calculate $v_i$, the initial velocity of the electrons. For an 
anode-cathode 
voltage of $V_{AC}=1000V$ for the electron gun, this gives 
$c/v_i=16.0077$, whereas the nonrelativistic expression gives 
$c/v_i=15.9843$. The positions in the Tables are very sensitive to 
variations in $c/v_i$.}.
 It is seen that in all cases the prediction of our scheme for the
position of the electron beam on the screen is distinctly different 
from the
Newtonian and Reissner-Nordstr\"{o}m (RN) 
predictions. One may be curious as to why the dispersion (see the last 
columns in Tables 1 and 2), $r_N-r_{MGR}$, between the Newtonian and 
Modified General Relativistic trajectories decreases as the potential 
$V(R)$ of the sphere increases. For weak potentials the curvature of
spacetime is small and the angle between the two trajectories is large,
as a result of which the two trajectories disperse more from each other
at large distances from the sphere
\footnote{The same phenomenon occurs in gravity between the Newtonian and
general relativistic trajectories of a neutral test particle moving in
the gravitational (Schwartzschild) field of a spherical mass. Replace
$m_E$ with $m_G$ in equations (24) and (28) to get the gravitational
equations.}. 
Therefore, by measuring the position of the electron beam on 
the screen the correct theory can be distinguished. In Figures 1-4, the 
trajectory of the electrons is drawn according to the three theories, where
again the differences in the trajectories are seen with certainty. For an 
anticipated difference of about 3-5 cm between $r_N$ and $r_{MGR}$, a 
rectangular metallic
box with dimensions $130cm\times 30cm\times 30cm$ with a circular lid near
the top of one end, and a glass window on the side facing the screen may be 
built very easily
\footnote{If evacuating the box is not a problem, a longer box can be
built to obtain larger $r_N-r_{MGR}$ (see Table 1). Figures 3 and 4,
on the other hand, suggest that a much smaller box could be used
for very large V(R) and anode-cathode voltage for the electron gun. 
Mathematically this is true. But the distances and angles must then 
be determined with perfect precision.}.
A rotary-diffusion pump system can easily obtain the
desired vacuum required for the electron gun to work. Care must be taken
to set the angles and the distances as precisely as possible because the
solutions are very sensitive to variations in them.

One may wonder, if in scattering experiments of the Rutherford type a 
deviation in the cross-section should have been seen due to the electrical
curvature of the spacetime. For the scattering of $\alpha$ particles
off gold nuclei, the correction term $2(q/m)_{\alpha}
k_eQ_{Gold}/(c^2r)=1.2\times 10^{-16}/r$ to $g^{MGR}_{00}$ turns out to be 
between $10^{-3}$ and $10^{-16}$ for $10^{-13}m\leq r\leq 1m$, where
$r$ is the position of the alpha-particle from the target nucleus.
So, within the precision of these experiments, no deviation from the 
cross-section can be seen.
\begin{table}
\label{Table 1}
\caption{Predicted positions, according to the three theories, of the 
electron beam at $\phi=5\pi/2$ after deflected by
a sphere of $R=5cm$ and potential $V(R)$ for an anode-cathode voltage
difference of $1000V$ for the electron gun located at $\phi=3\pi/2$ and
$r_i=15cm$ from the center of the sphere.}
\vspace{0.5cm}
\begin{tabular}{|c|c|c|c|c|c|c|} \hline
V(R) & $\frac{m_E}{R}$ & $\alpha$  & $r_{RN}$ & $r_N$ & $r_{MGR}$ & 
$r_N-r_{MGR}$\\
(Volt) &  &(degree)  & (cm) & (cm) & (cm) & (cm)\\ \cline{1-7}
1325 & $2.59\times 10^{-3}$ & 40  & 212.08 & 208.07 & 190.71 & 17.36\\
1350 & $2.64\times 10^{-3}$ & 40  & 165.02 & 162.48 & 151.51 & 10.97\\
1375 & $2.69\times 10^{-3}$ & 40  & 135.06 & 133.27 & 125.68 &  7.59\\
1400 & $2.74\times 10^{-3}$ & 40  & 114.31 & 112.97 & 107.37 &  5.60\\
1425 & $2.79\times 10^{-3}$ & 40  &  99.08 &  98.03 &  93.72 &  4.31\\
1600 & $3.13\times 10^{-3}$ & 45  & 219.53 & 214.91 & 198.36 & 16.55\\
1625 & $3.18\times 10^{-3}$ & 45  & 176.50 & 173.38 & 162.30 & 11.08\\
1650 & $3.23\times 10^{-3}$ & 45  & 147.57 & 145.31 & 137.34 &  7.97\\
1700 & $3.33\times 10^{-3}$ & 45  & 111.15 & 109.76 & 105.03 &  4.73\\
1750 & $3.42\times 10^{-3}$ & 45  &  89.15 &  88.19 &  85.03 &  3.16\\
1875 & $3.67\times 10^{-3}$ & 50  & 225.14 & 220.00 & 204.86 & 15.14\\
1900 & $3.72\times 10^{-3}$ & 50  & 185.61 & 181.99 & 171.39 & 10.60\\
1925 & $3.77\times 10^{-3}$ & 50  & 157.88 & 155.17 & 147.32 &  7.85\\
1950 & $3.82\times 10^{-3}$ & 50  & 137.37 & 135.25 & 129.18 &  6.07\\
2000 & $3.91\times 10^{-3}$ & 50  & 109.03 & 107.61 & 103.66 &  3.95\\ 
\hline
\end{tabular}
\end{table}
\begin{table}
\label{Table 2}
\caption{Same as Table 1, but $R=2.5cm$ and $r_i=10cm$.}\vspace{0.5cm}
\begin{tabular}{|c|c|c|c|c|c|c|} \hline
V(R) & $\frac{m_E}{R}$ & $\alpha$  & $r_{RN}$ & $r_N$ & $r_{MGR}$ & 
$r_N-r_{MGR}$\\
(Volt) &  &(degree)  & (cm) & (cm) & (cm) & (cm)\\ \cline{1-7}
2100 & $4.11\times 10^{-3}$ & 45 & 193.55 & 188.39 & 170.01 & 18.38\\
2125 & $4.16\times 10^{-3}$ & 45 & 155.86 & 152.40 & 140.01 & 12.38\\
2150 & $4.21\times 10^{-3}$ & 45 & 130.45 & 127.95 & 119.02 &  8.93\\
2200 & $4.31\times 10^{-3}$ & 45 &  98.38 &  96.87 &  91.56 &  5.31\\
2250 & $4.40\times 10^{-3}$ & 45 &  78.97 &  77.94 &  74.39 &  3.55\\
\hline
\end{tabular}
\end{table}
\begin{figure}
\epsfig{figure=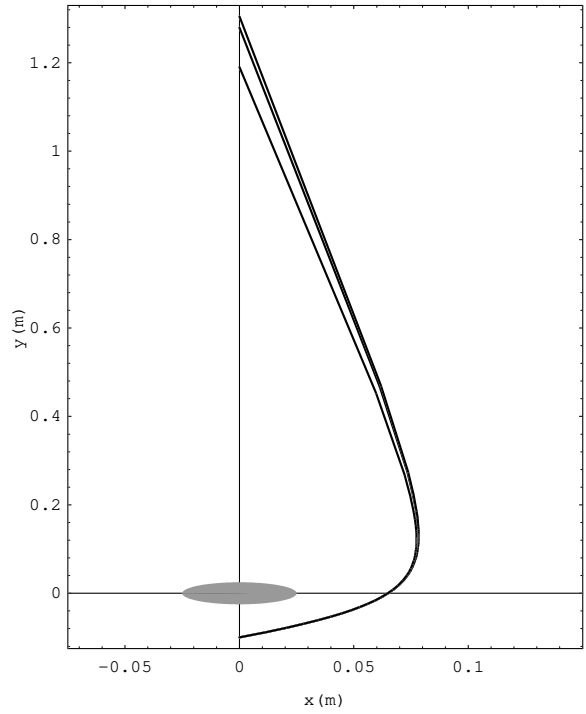, width=14cm}
\caption{ The trajectories of the electron beam according to the
Reissner-Nordstr{\o}m (the top curve), Newtonian (the middle curve), and
the Modified General Relativity (the bottom curve) theories for an
anode-cathode voltage of $1000V$ for the electron gun, and for a sphere 
of $R=2.5cm$ and $V(R)=2250V$.
\label{fig1}
}
\end{figure}
\begin{figure}
\epsfig{figure=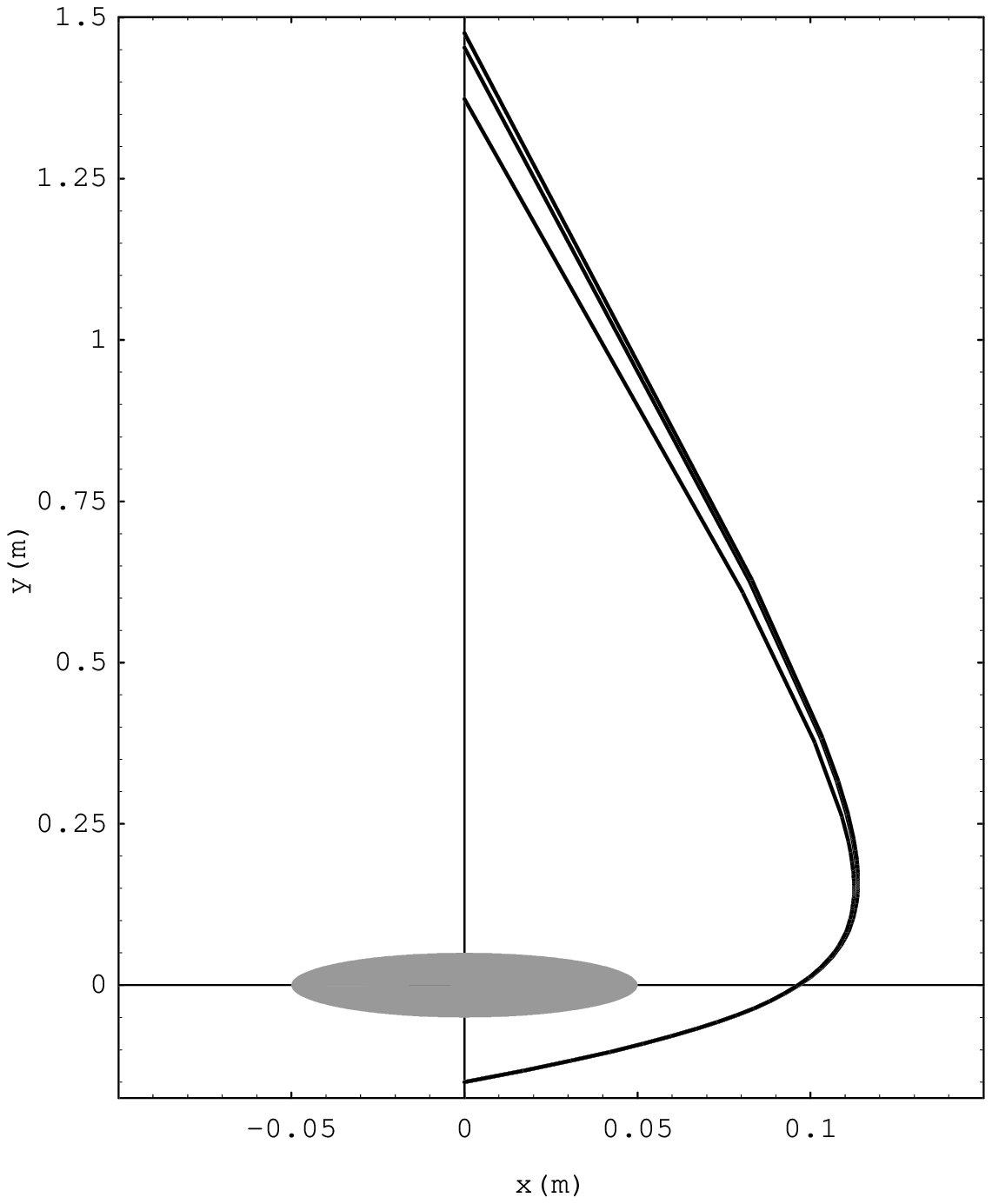, width=14cm}
\caption{Same as Figure 1, but $R=5cm$ and $V(R)=1750V$.
\label{fig2}
}
\end{figure}
\begin{figure}
\epsfig{figure=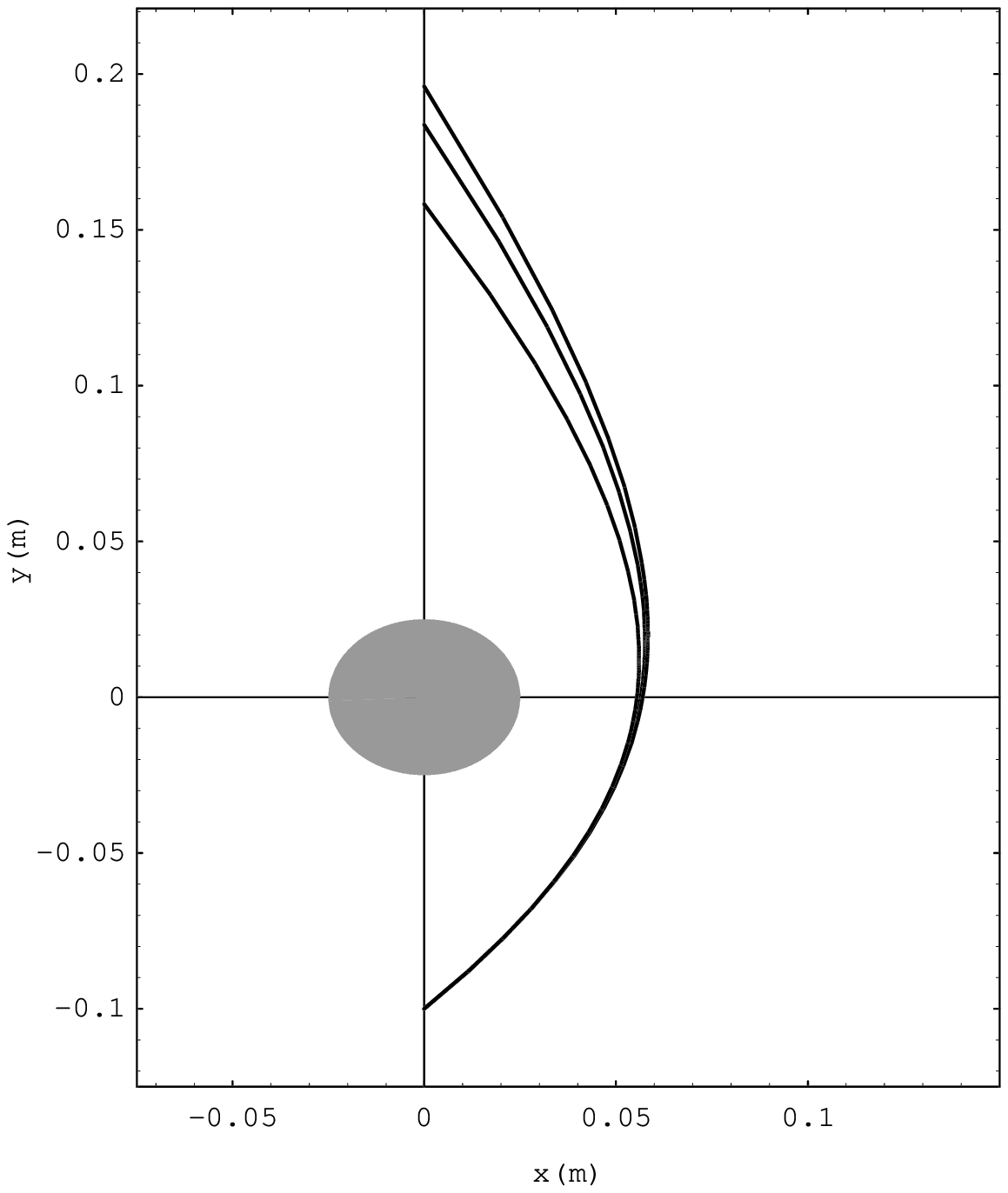, width=14cm}
\caption{Same as Figure 1, but the anode-cathode voltage for the
electron gun is $10000V$, $R=2.5cm$ and $V(R)=30000V$.
\label{fig3}
}
\end{figure}
\begin{figure}
\epsfig{figure=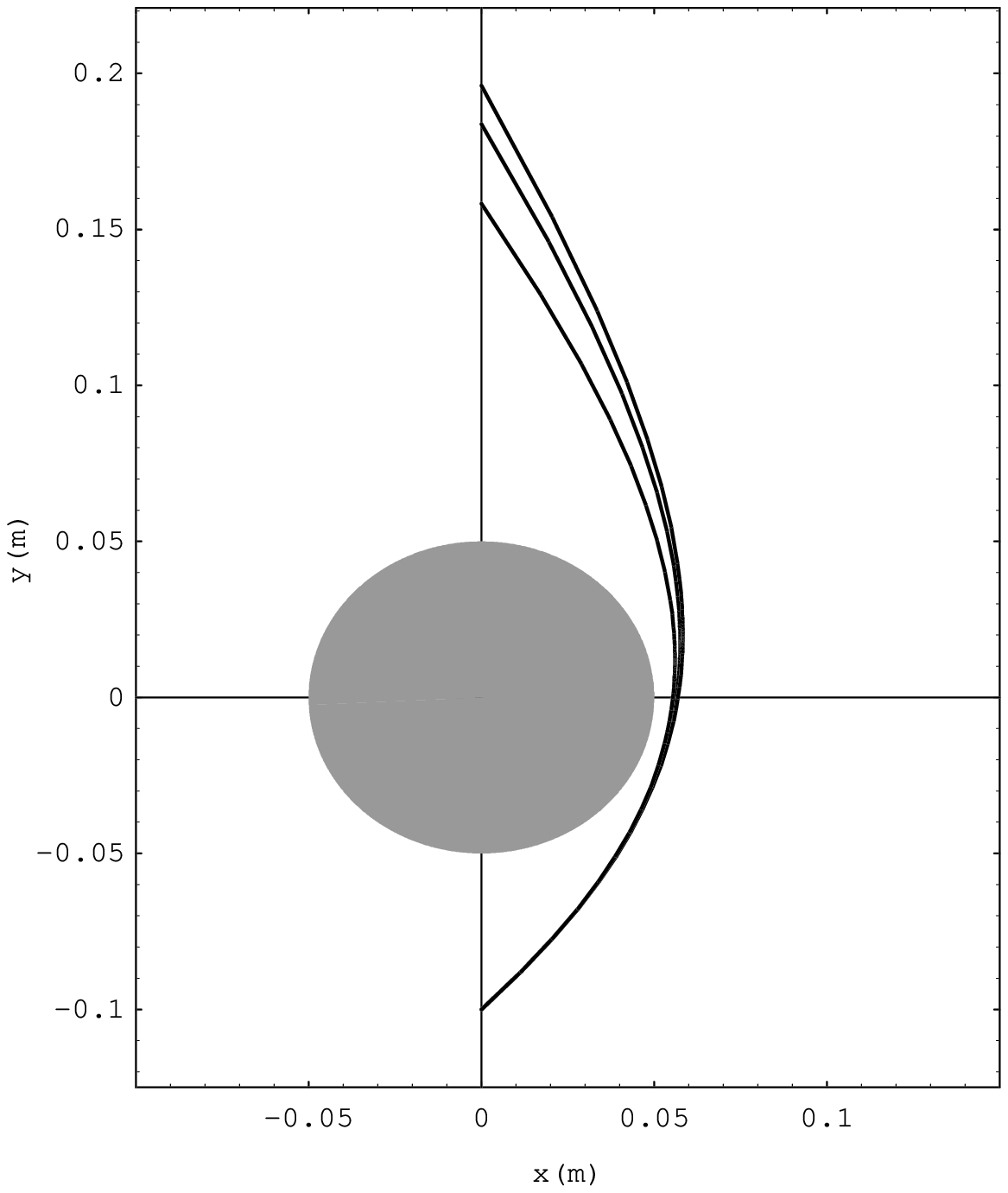, width=14cm}
\caption{Same as Figure 1, but the anode-cathode voltage for the electron
gun is $10000V$, $R=5cm$, and $V(R)=15000V$.
\label{fig4}
}
\end{figure}

{\bf II. The Electrical Redshift of Light:} If true, one immediate and 
dramatic consequence of the gravito-electromagnetic unified 
description in our
scheme is that light should undergo a redshift as it travels against a
uniform electric field. The existence of the electrical redshift can be 
inferred from the eqivalence principle for the electric field [1]. Consider
a cabin and two clocks seperated by a horizontal distance $d$ in it, all
with the same $q/m$ ratio. For definiteness, assume the charges are positive.
Let the cabin be accelerating to the left at the rate $a=(q/m)E$ to
simulate an electric field $E$ directed to the right. An inertial observer 
describes the following
chain of events: The right and left-hand clocks are both accelerating
to the left with acceleration $a$. The right-hand clock is sending photons
to the left-hand clock at the rate $\nu_R$ photons per second. It takes
time $t=d/c$ for a photon to reach the left-hand clock, during which time
the velocity of the left-hand clock increases by $\Delta v=(q/m)Ed/c$.
Therefore the rate $\nu_L$ that the photons are detected by the left-hand
clock is decreased by a Doppler redshift
\begin{equation}
\nu_L=\nu_R\left(1-\frac{\Delta v}{c}\right)=\nu_R\left(1-\frac{q}{m}
\frac{Ed}{c^2}\right).
\end{equation}
This means that the frequency of a photon detected by the left-hand clock
undergoes a Doppler shift exactly as in eq.(33). Therefore the fractional
change in the frequency of the photons is
\begin{equation}
\frac{\Delta\nu}{\nu}=\frac{\nu_L-\nu_R}{\nu_R}=-\frac{q}{m}\frac{Ed}{c^2},
\end{equation}
where now $\nu$ refers to the photon frequency. Then according to the
eqivalence principle, the same redshift must be observed as light travels
to the left in a uniform static electric field $E$ directed to the right. 
Note, strange as it may sound though, that the above argument implies
that the photon behaves in an electric field as if it has a nonzero 
 ``effective electric charge'' 
and hence an electric charge-to-mass ratio $(q/m)_{\gamma}$
\footnote{This is similar to the gravitational situation in which the
photon has ``effective'' gravitational and inertial masses and 
$(m_g/m_i)_{\gamma}=1$. In the electrical case, however,  we do not know
the value of $(q/m_i)_{\gamma}$. It must be determined from the 
experiment.}
\footnote{Having found out that photons have a nonzero electric-charge-
to mass ratio, we point out that the cabin and the clocks then must
have this very same ratio so that the equivalence principle for the
electric field is applicable. However, one should not conclude
from this that, in reality the atoms (clocks) emitting and absorbing the 
photons must have the same electric charge-to-mass ratio as the photons. 
This can be seen by excluding the clocks from the cabin in the
above thought experiment, or from the conservation of energy argument
as applied to a particle moving in a uniform electric field and
converting to a photon. This argument does not involve any ``clocks''.}.
Note, however, that the above argument does not fix the sign of 
the effective charge of the photon. If the effective charge is
negative, photons then would be redshifted as they moved in the 
same direction as the electric field. Hence, assuming a 
positive ``effective electric charge'' for
the photon, the conservation of energy of a particle moving in an electric 
field and then converting to a photon, just like a particle falling
in a gravitational field and then converting to a photon [7], yields the 
same redshift as in eq.(34).

An experiment of the Pound-Rebka-Snider type [8,9] can be done to verify
the redshift and/or to put a limit on the $(q/m)_{\gamma}$  of the photon. 
A $\Delta\lambda/\lambda$ of $\approx 10^{-15}$ should be seen for a voltage
difference of about $100V$ between the detection and emission points of
the photons if $(q/m)_{\gamma}=1C/kg$
\footnote{Note, as we have pointed out in [1], that in a different
system of units the electric charge $q$ and the mass $m$ may be measured
in the same unit. In such a system of units, $(q/m)_{\gamma}=\pm 0/0
\rightarrow \pm 1$ seems  more likely.}. If, on the other 
hand, $(q/m)_{\gamma}=0.1C/kg$ or $0.01C/kg$, the required voltage 
difference would be about $10^3V$ or $10^4V$, respectively.

Before we end this section, we would like to remark that a nonzero
$(q/m)_{\gamma}$ implies that light would be deflected or 
scattered off as it passess
a charged spherical object just as it is deflected by a massive spherical
object like the sun. The magnitude of the deflection, however, is so small,
 even for $(q/m)_{\gamma}=1C/kg$, that a laboratory experiment does not 
seem possible.

{\bf III. The Deflection of Light in a Magnetic Field:} Another consequence 
of a nonzero elctric charge-to mass ratio for the photon is that light
would be deflected in a magnetic field. Consider a uniform static magnetic
field $B$ directed downward in the $-z$ direction. Let a light beam be
emitted from a point and travel in the $xy$ plane so that the velocity
of the light beam is perpendicular to the magnetic field. The light beam
should travel in a counterclockwise circle of radius
\begin{equation}
R=\frac{1}{(q/m)_{\gamma}}\frac{c}{B},
\end{equation}
which follows from the equality of the centripetal and magnetic forces
on a single photon. Let $d$ be a straight distance that a photon would have
travelled had it been not deflected by the magnetic field. Then the
deflection $\Delta$, the distance from the end of the distance $d$ to the
actual position of the photon on the circle, is
\begin{equation}
\Delta =\frac{1}{(q/m)_{\gamma}}\frac{c}{B}\left\{1-cos\left[sin^{-1}
\left(\left(\frac{q}{m}\right)_{\gamma}\frac{Bd}{c}\right)\right]\right\}.
\end{equation}
Tabulated in Table 3 are the deflections for $(q/m)_{\gamma}=1C/kg$ a light
beam would suffer as a function of $B$ and the straight distance $d$, 
the distance light is allowed to travel when $B=0$. We see that a deflection
of a tenth of a millimeter is expected for $B=1T$ and $d=250m$. A uniform
magnetic field extending to a desired length can easily be obtained by
placing a number of electromagnets end-to-end. The positions of a light beam
on a ``film'' in the absence and presence of the magnetic field can be
measured. The distance between the two positions would be the anticipated
deflection.

In this letter, we have proposed three experiments to test whether or not 
gravitation and electromagnetism have a unified description
through a symmetric metric tensor. The  experiment of the deflection of an 
electron beam by
a positively charged sphere, which is to show if a distribution of
electric charge  curves the spacetime independently of its 
gravitational field, is the simplest one and shoud be done first. The other
two experiments depend strongly on the predicted electric-charge-to-mass
ratio for the photon. A negative result in these experiments would
still be useful to place an upper limit on  $(q/m)_{\gamma}$. 
\begin{table}
\label{Table 3}
\caption{The deflections, $\Delta$, a light beam is expected to suffer
for $(q/m)_{\gamma}=1C/kg$ as a function of the magnetic field $B$ and the
distance $d$ the beam travels when $B=0$.}\vspace{0.5cm}
\begin{tabular}{|c|c|c|} \hline
B & d & $\Delta$  \\
(T) &  (m)  & (mm) \\ \cline{1-3}
1 & 250  & 0.104\\
1 & 500  & 0.417 \\
1 & 1000 & 1.668 \\
5 & 100  & 0.083\\
5 & 150  & 0.188\\
5 & 250  & 0.521\\
5 & 350  & 1.022\\
10 & 50  & 0.042\\
10 & 100 & 0.167\\
10 & 250 & 1.042\\
\hline
\end{tabular}
\end{table}\vspace{0.2cm}

\noindent{\bf Acknowledgements}\vspace{0.2cm}

We are grateful to Prof. Mahjoob O. Taha for invaluable 
discussions. We thank Mr C\"{u}neyt Elibol, Dr Orhan \"{O}zhan
, and Dr Arif Akhundov for various comments.\vspace{0.2cm}

\noindent{\bf References}\vspace{0.5cm}

\noindent {[1]} M. \"Ozer, On the Equivalence Principle and a Unified
Description of Gravitation and Electromagnetism, gr-qc/9910062.\\
{[2]} A. Einstein, Ann. d. Phys. 49(1916)769.\\ 
{[3]} H. Reissner, Ann. d. Phys. 50(1916)106.\\
{[4]} G. Nordstr{\o}m, Proc. Kon. Ned. Akad. Wet. 20(1918)1238.\\
{[5]} C. W. Misner, K. S. Thorne, and J. A. Wheeler, Gravitation,
      W. H. Freeman and Company, 1973.\\
{[6]} K. Schwarzschild, Berl. Ber. (1916)189.\\
{[7]} See, for example, ref.[5], p.187.\\
{[8]} R. V. Pound and G. A. Rebka, Phys. Rev. Lett., 4(1960)337.\\
{[9]} R. V. Pound and J. L. Snider, Phys. Rev., B 140(1965)788.

\end{document}